\documentclass[summary]{ursi}
\usepackage{tabularx}
\usepackage{graphicx}
\usepackage{txfonts}
\usepackage{hyperref}

\def\HI{\ion{H}{I}}

\usepackage{printlen}
\usepackage{layouts}

\newcommand{\kms}{$\,$km~s$^{-1}$}

\newcommand{\msun}{{${\rm M}_\odot$}}

\newcommand{\cmsq}{cm$^{-2}$}

\newcommand{\eg}{\mbox{e.g.}}
\newcommand{\ie}{\mbox{i.e.}}

\newcommand{\meer}{{MeerKAT}}

\newcommand{\nhi}{{$N_{\rm H {\hskip 0.02cm \tt I}}$}}
\newcommand{\mhi}{{$M_{\rm H {\hskip 0.02cm \tt I}}$}}

\title{Neutral Atomic Hydrogen surveys: past, present and future}

\author{F.~M.~Maccagni$^*$\affref{ref1},  W.~J.~G. de Blok\affref{ref2}\affref{ref3}\affref{ref4}}
\affiliation{%
  \aff{ref1}{INAF -- Osservatorio Astronomico di Cagliari, via della Scienza 5, 09047, Selargius (CA), Italy}
  \aff{ref2}{ASTRON -- Netherlands Institute for Radio Astronomy, Oude Hoogeveensedijk 4, 7991 PD, Dwingeloo, The Netherlands}
  \aff{ref3}{Dept. of Astronomy, Univ. of Cape Town, Private Bag X3, Rondebosch 7701, South Africa}
  \aff{ref4}{Kapteyn Astronomical Institute, University of Groningen, PO Box 800, 9700 AV Groningen, The Netherlands}
}

\begin{document}

\maketitle

\begin{abstract}
 Neutral atomic hydrogen (\HI) observations are fundamental to understand the dynamics of galaxies, their assembly, the fuelling of their star formation and environmental interactions. \HI\ studies have so far been limited by the capabilities of single-dish radio telescopes or synthesis arrays to either small samples or low resolution and sensitivities. Now, the Square Kilometer Array precursors and pathfinders are providing a novel view of the \HI\ in and around galaxies allowing wide-field high resolution deep surveys in nearby galaxies. We give an overview of past, current and future \HI\ surveys consistently comparing their \HI\ column density and spatial resolutions highlighting their main scientific key goals and results.  
\end{abstract}

\section{Neutral hydrogen gas in galaxies}

Over cosmic time, star formation is one of the main drivers of galaxy evolution. It creates new stars, while simultaneously consuming the cold gas fuel reservoir. The availability of the latter is determined by the delicate balance between the amount of gas consumed by star formation, the amount of material accreted from the inter-galactic medium (IGM) and the amount of gas expelled from the galaxy. In spiral galaxies such as our own, to sustain SF over billions of years the gas reservoirs must be replenished from the IGM through cosmic time. In elliptical galaxies SF has been rapidly quenched and the gas reservoirs have been depleted and kept so for billions of years. Several phenomena determine the fate of a galaxy. In the environment, merger-driven tidal motions between galaxies or hydrodynamic interactions can drive gas into galaxies or strip it from them. Cold gas infall from the inter-galactic medium and collapse into dense clouds triggers star formation, whose supernovae and stellar winds may heat and expel the gas (\ie\ `baryon-cycle). Cold gas may be funnelled down to the central regions triggering nuclear activity, whose energetic feedback can also eject gas from the galaxy and contribute to quench its star formation.

Hydrogen is the most abundant element in the Universe and is the main component of the gas fuel reservoir from which galaxies are replenished. Sufficiently cold hydrogen can be observed at 21 cm and therefore can be potentially used to trace these gas reservoirs. This neutral atomic hydrogen (HI) also forms the main component of the gas inside disk and dwarf galaxies. A complete knowledge of the \HI\ content in galaxies (\eg\ the \HI\ mass function between $\sim10{^6}$ and $\sim 10^{11}$~\msun) and how this correlates with their other gas and stellar properties is crucial to understand how galaxies sustain their star formation.
\HI\ gas typically extends beyond the stellar disk, this has allowed to probe the rotation curves of galaxies. For many galaxies, these are flat which implies a divergence between the dynamics expected by the visible matter and the total matter distribution~\cite{bosma}. \HI\ is thus an exquisite tracer to study the Dark Matter content in galaxies and identify `exotic' sources which may deviate from the typical assembly expected by the $\Lambda$CDM cosmological model,~\eg~\cite{pavel}.

\HI\ emits through its hyper-fine transition at 21-cm wavelengths and is observable in the nearby Universe by radio telescopes in the L-band. For the reasons described above, \HI\ surveys have always been key science projects of both single dish and radio synthesis telescopes. The development of new facilities and sensitive receivers has pushed the investigation of cold gas in galaxies to previously unexplored low column density sensitivities (\nhi$\lesssim 10^{19}$~\cmsq), high angular resolutions (10-30'') and over wide areas of the sky ($\gtrsim 20$ deg$^2$), thus allowing us to obtain a detailed view of galaxy dynamics and assembly over large statistical samples of galaxies. In the following sections, we compare the different areas, sensitivities and resolutions achieved by past and on-going \HI\ surveys and will show the potentials of the upcoming Square Kilometer Array (SKA~\cite{ska}) and Deep Synoptic Array (DSA2000)~\cite{dsa2000} telescopes.

\section{Past \HI\ surveys}
\label{sec:past}
\begin{figure}[htbp]
  \centering
    \includegraphics[width=80mm]{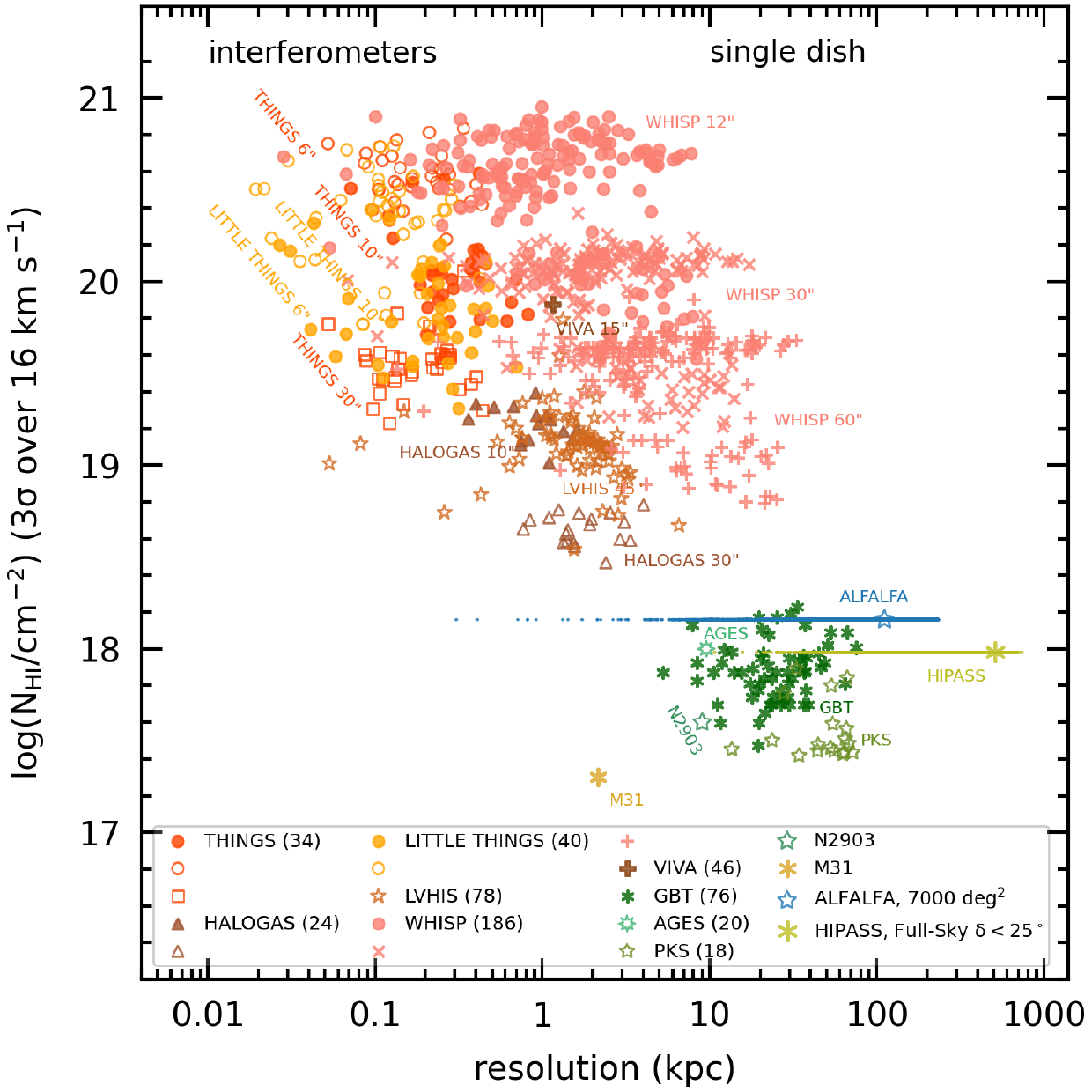}
  \caption{\HI\ column density sensitivity versus spatial resolution of past \HI\ surveys. Symbols show the $3\sigma$ column density sensitivity over $16$\kms, for different interferometric and single dish surveys (see Sect.~\ref{sec:past} for further details). The angular resolution of the observations has been converted to spatial resolution according to the distance of the targets (see text for more details). The sample sizes are shown by the numbers within parenthesis in the legend.}
  \label{fig:past}
\end{figure}

Neutral hydrogen studies are always limited by the capacities of the available radio telescopes. Single-dish telescopes (\ie\ Parkes, Arecibo, Green Bank Telescope) guarantee a wide-field of view and high surface brightness sensitivity but have arcminute scale angular resolution, while radio synthesis telescopes  (\ie\ ATCA, VLA, WSRT) enable arcsecond scale resolution observations over smaller fields of view ($\lesssim 1$ deg$^2$). The incomplete coverage of the \emph{uv}-plane due to the limited number of antennas limits the surface brightness sensitivity to the faint extended emission of \HI\ in the outskirts of galaxies and in the IGM. In Figure~\ref{fig:past} we compare the column density sensitivity and physical scale resolution of several surveys carried out with single dish or interferometers. All sensitivities are homogenised to a $3\sigma$ limit over a velocity range of $16$~\kms. This velocity is representative of the lowest velocity dispersions commonly observed in galactic disks. We convert the angular resolution of each survey based on the distances of the targets. The Figure shows that so far \HI\ studies have focused on two separate regions of parameter spaces. Single-dish telescopes have focused on sensitive ($\sim 10^{18}$~\cmsq) observations over large portions of the sky, but mostly rely on unresolved detections (resolution $\gtrsim 10$ kpc). Interferometers, instead, have focused on highly resolved (resolution $\lesssim 1$ kpc) observations of a limited number of galaxies in the Local Universe, providing the most detailed studies of dynamics in galaxies from dwarfs to spirals. The surveys shown are the Westerbork observations of neutral Hydrogen in Irregular and SPiral galaxies (W\HI SP,~\cite{whisp}), the Local Volume HI Survey (LV\HI S, ~\cite{lvhis}), the \HI\ Nearby Galaxy Survey (T\HI NGS,~\cite{things}), the Local Irregulars That Trace Luminosity Extremes THINGS (LITTLE T\HI NGS,~\cite{littlethings}), Hydrogen Accretion in LOcal GAlaxieS (HALOGAS,~\cite{halogas}) and the VLA Imaging of Virgo in Atomic Gas (VIVA,~\cite{viva}). For single dish studies we show the full samples of The H I Parkes All Sky Survey (HIPASS,~\cite{hipass}) and the Arecibo Legacy Fast ALFA (ALFALFA,~\cite{alfaalfa}). For these two surveys, the markers show the median redshifts of the detections (40 and 110 Mpc respectively). With these surveys it has been possible to obtain a near-complete census of the overall \HI\ density in the nearby Universe and of the \HI\ mass function. It has also provided a comprehensive picture of the various scaling relations between the neutral atomic phase and the gas and stellar content in galaxies. The high sensitivity of single dish telescopes has also enabled targeted deep observations in search for the diffuse cold clouds and filaments that may accrete onto galaxies (AGES~\cite{ages}) and the observations of \mbox{NGC~2903} and M31~\cite{n2903,m31} and GBT and Parkes observations of nearby galaxies~\cite{gbt1,gbt2,gbt3}. However, when diffuse gas has been detected, the question whether it was accreting onto the galaxy or was a remnant of an interaction has remained an unanswered question, due to the limited angular resolution of the observations.

The Figure also highlights that interferometers provide a continuous description of the angular-resolution column-density sensitivity parameter space. By changing the weights of the short or long baselines (\ie\ robustness) one can trade angular resolution for sensitivity thus enabling the detection of diffuse and extended \HI\ features in the outskirts and environment of galaxies. Nevertheless, reaching the column densities probed by single-dish observations for a representative number of sources is almost impossible using realistic observing times.

\begin{figure*}[htbp]
  \centering
    \includegraphics[width=80mm]{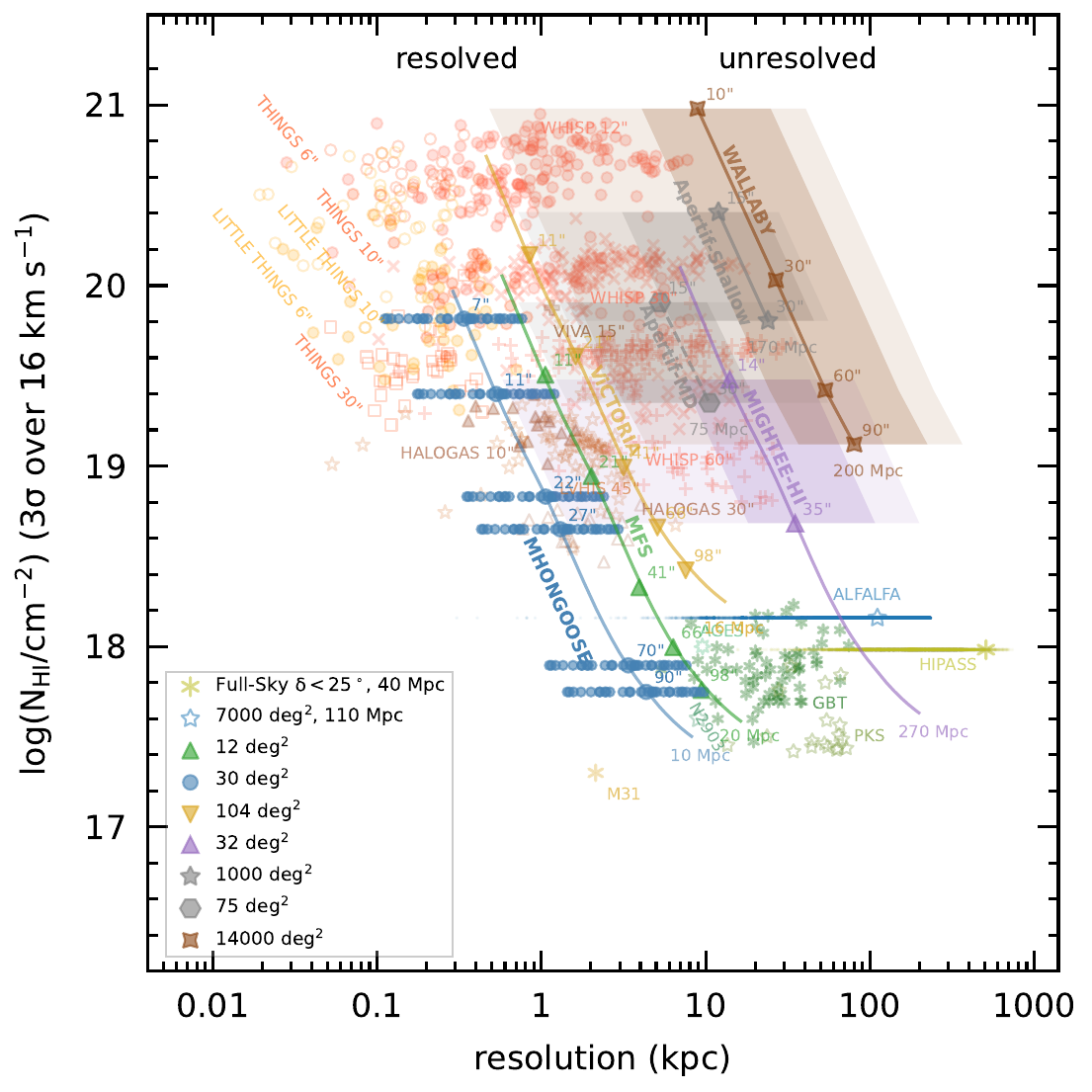}
    \includegraphics[width=80mm]{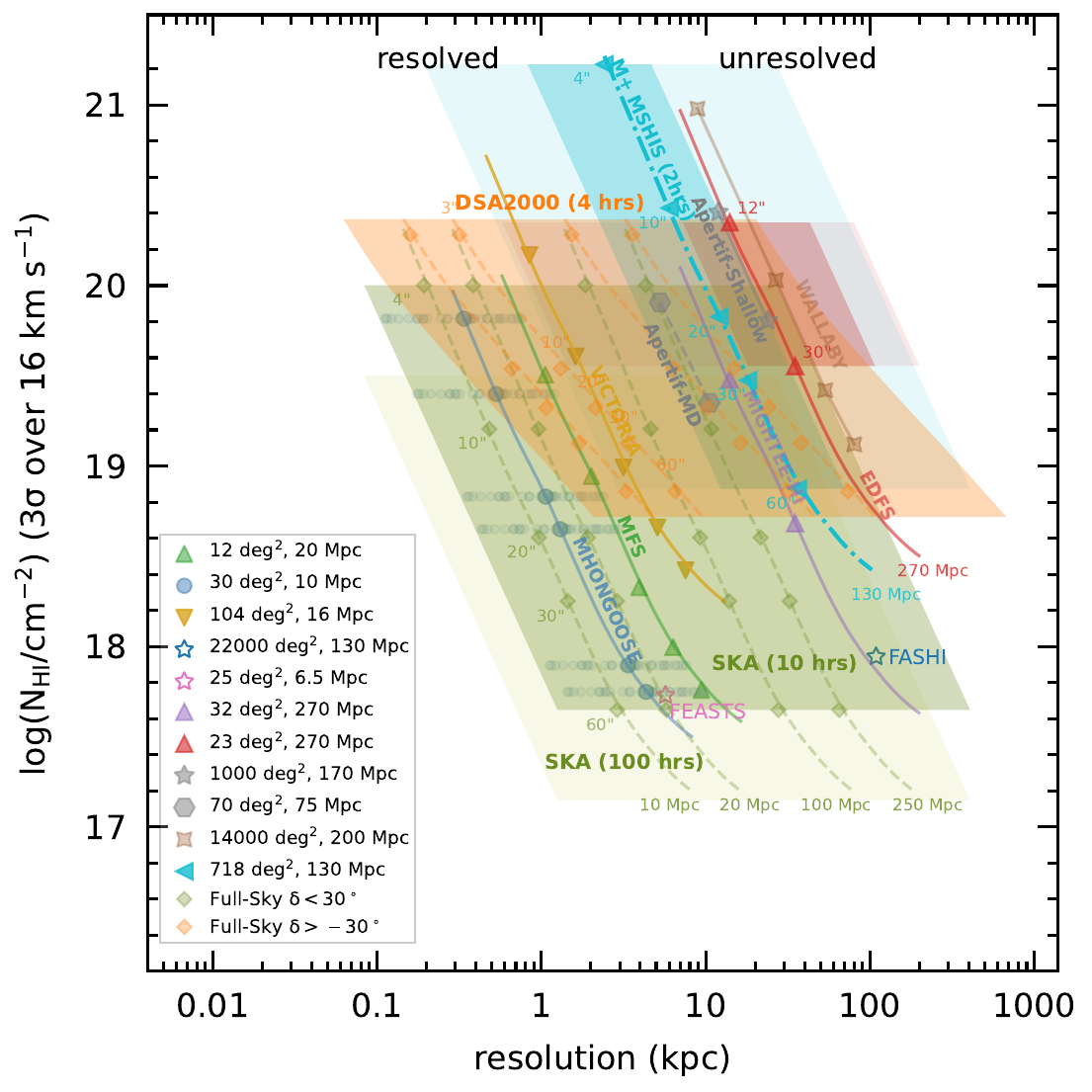}
  \caption{{\em Left}: Sensitivity versus resolution of past and on-going \HI\ surveys. For wide-field surveys, the lines show the median distance of the expected \HI\ detections of the surveys, the dark shaded areas the resolutions reached for 90$\%$ of the expected detections, and the light shaded areas the resolutions between 10 Mpc and the maximum accessible redshift of the survey. {\em Right}: Sensitivity versus resolution of on-going and future surveys. Lines and shadings are as in the left panel.}
  \label{fig:current}
\end{figure*}

\section{Current \HI\ surveys}
\label{sec:current}

In the last decade, the SKA L-band precursors and pathfinders (MeerKAT, ASKAP and Apertif) opened a new parameter space for \HI\ interferometric observations. For example, MeerKAT~\cite{meerkat} can provide wide-field of view ($\sim 1$ deg$^2$ and/or high surface brightness sensitivity (\nhi$\gtrsim 10^{18}$~\cmsq), with $10''$ with a few tens of hours of integration time. This  corresponds to an \HI\ mass sensitivity of $\sim 10^6$~\msun for an unresolved source at 10~Mpc assuming a linewidth of $50$~\kms. The new phased array feed receivers of WSRT (Apertif,~\cite{apertif1}) and ASKAP~\cite{askap} have increased the field of view of interferometers by an order of magnitude, thus enabling blind interferometric surveys over several thousands of square degrees, also increasing the studies of high redshift \HI\ sources ($\gtrsim 100$ Mpc). Figure~\ref{fig:current} (left) compares the column density sensitivity of SKA precursors and pathfinders \HI\ surveys with the past \HI\ surveys presented in the previous Section. The phased array feed telescope surveys shown are the Widefield ASKAP L-band Legacy All-sky Blind surveY (WALLABY,~\cite{wallaby1}) and the Apertif-Shallow and Medium-Deep surveys~\cite{apertif2}. \meer\ surveys are the MeerKAT International GHz Tiered Extragalactic Exploration in HI (MIGHTEE-HI,~\cite{maddox}), the MeerKAT Fornax Survey (MFS,~\cite{mfs}), the MeerKAT \HI\ Observations of Nearby Galactic Objecs: Observing Southern Emitters (MHONGOOSE,~\cite{mhon}) and the Virgo Cluster multi Telescope Observations in Radio of Interacting galaxies and AGN (VicTORIA,~\cite{boselli}). For wide-field surveys (WALLABY, Apertif and MIGHTEE-HI) the lines show the spatial resolution at the median distance of the expected \HI\ detections. The dark shaded regions show the range of resolutions for $90\%$ of the detections and the light shaded regions the resolutions for the full sample. The Figure shows that WALLABY and Apertif are wide-field surveys with sensitivities and resolutions that previously have been limited only to targeted observations in the Local Volume (LVHIS and WHISP). With resolutions between 5 and 100 kpc these surveys will provide the most complete census of \HI\ in galaxies at distances ($\gtrsim 70$ Mpc), which so far has mostly been studied only with single-dish surveys. Nevertheless, the system temperature of phased array receivers does not allow to easily reach the very low \HI\ column densities needed to properly detect and characterize, for example, cold gas accretion features. These studies are best tackled by the \meer\ large survey programs which, for the very first time, are reaching single-dish column density sensitivity with ten times higher angular resolution. The MeerKAT Fornax Survey is observing the nearby Fornax cluster and infalling group of Fornax A~\cite{mfs}. MHONGOOSE is targeting 30 nearby galaxies from dwarfs to star forming spirals to detect and characterise the diffuse \HI\ that is potentially accreting onto galaxies and understand its connection to SF. ViCTORIA is observing the entire Virgo cluster that was previously only observed with targeted observations of 46 galaxies with VIVA. MIGHTEE-HI is blindly targeting four fields to provide a sensitive unbiased characterisation of the \HI\ population at $250$ Mpc. 

The Figure shows that even if investigating a new parameter space in sensitivity, resolution and area, the current interferometric \HI\ surveys are still limited by the different optimizations of their telescopes. Phased array feed telescopes and \meer\ provide similar $10-90$ arcsecond angular resolutions, but while the first are best suited for wide-field surveys, they do not reach extremely low column densities. \meer\ enables these deep studies but over only limited fields, samples of galaxies and environments. Table~\ref{tab:surveys} summarises the main properties of the on-going \HI\ surveys. 

\section{Future \HI\ surveys}
\label{sec:future}

In the next few years, new radio telescopes will keep exploring new regions of the sensitivity versus resolution parameter space allowing deep wide-field \HI\ surveys with improved angular resolution. The upgrade from \meer\ to \meer+ will increase the survey speed of the instrument and provide higher (4'') angular resolution. This motivates the development of a medium-shallow \HI\ survey (M+MSHIS) which, compared to current \meer\ \HI\ studies, will investigate over a wide area (718 deg$^2$), reaching with only 2 hours per pointing, \HI\ sensitivities $\lesssim 10^{20}$~\cmsq. The sensitivity versus angular resolution of the M+MSHIS survey is shown in Figure~\ref{fig:current} (right) along with the on-going \HI\ surveys. Before this upgrade, as a pilot for a medium-shallow survey, \meer\ is observing the Euclid Deep Field South (EDFS). This is an example of the synergy between \meer\ and deep optical observations (from the \emph{Euclid} space mission), which enable the unambiguous identification of low-surface brightness gas rich galaxies.
The Figure shows that an unprecedented and complete view of the \HI\ in and around galaxies will be provided by DSA~2000 and the SKA observations. Thanks to 2000 antennas guaranteeing a dense coverage of baselines within 10 km, DSA 2000 will reach M+MSHIS sensitivities at 20 arcseconds in half the integration time. Because of the survey strategy of the instrument, DSA 2000 will scan the full northern sky, thus providing the most complete sample of resolved \HI\ sources in the Northern Hemisphere. Besides survey speed, an important innovation of DSA 2000 is the improvement in angular resolution to 3''. The right panel of Fig.~\ref{fig:current} shows that with only 4 hours of integration DSA 2000 will detect the \HI\ in the star-forming disks (\nhi$\gtrsim 10^{20}$~\cmsq) at 100 pc resolution (double the resolution of MHONGOOSE). By increasing the sensitivity of one order of magnitude w.r.t. Apertif and WALLABY it will be possible to perform a direct comparison in all SF galaxies of the nearby universe (20000 sources from dwarfs to starburts within $100$ Mpc), between the \HI\ distribution and kinematics and the other gas phases of the ISM and the stars to probe how the baryon cycle within galaxies fuels star formation~\cite{dsa20002}.

The SKA \HI\ observations will be groundbreaking. They will reach between $10^{19}$ and $10^{18}$~\cmsq\ at unmatched resolutions (5-50) arcseconds in only 10 hours per pointing\footnote{Here we assume the SKA-MID baseline design with $197$ dishes.}. This will allow us to perform MFS and MHONGOOSE surveys not anymore on limited samples and environments but over the entire Southern Hemisphere (\ie\ hundreds of thousands of galaxies). SKA observations with 100 hours per pointing (light shaded green area in the figure) will potentially give a definitive answer to cold gas accretion in galaxies and on the existence of cold gas cosmic filaments: either a detection and full characterisation of these systems will be possible either a non-detection will require to improve our understanding of the physical processes that sustain star formation over cosmic time. The Figure also shows the on-going \HI\ survey of nearby galaxies with the single-dish FAST telescope (FEASTS,~\cite{feasts}) and the planned FAST-\HI\ wide-field survey FASHI,~\cite{fashi}. This survey will be complementary to \meer\ and SKA observations providing high \HI\ sensitivities on even larger angular scales.

The other game-changer of SKA \HI\ surveys is the complete description of the neutral gas phase in galaxies out to redshift (z$\sim 0.5-1$).  High redshift studies have so far been limited to `pencil beam' surveys in emission such as the Cosmos \HI\ Extragalactic Survey (CHILES,~\cite{chiles}), Looking At the Distant Universe with the MeerKAT Array (LADUMA,~\cite{laduma}) and the Deep Investigation of Neutral Gas Origins (DINGO,~\cite{dingo}), or to targeted \HI\ absorption surveys such as the search of associated absorbing systems in galactic nuclei (\eg~\cite{maccagni}) of 21-cm absorbers in damped Lyman-alpha systems towards radio quasars (\eg~\cite{kanekar}) and the on-going First Large Absorption Survey in HI (FLASH,~\cite{flash}) and the MeerKAT Absorption Line Survey (MALS,~\cite{mals}) or to stacking experiments (\eg~\cite{chowdury}). Large-volume \HI\ surveys of resolved sources will enable to characterise the \HI\ mass function throughout the last Gyr and the cosmic \HI\ density. This will allow us to understand if and how the strong decrease in star formation rate observed in the last billion years is linked to a variation in their \HI\ gas reservoir. 
 
 \begin{table*}[tbh]
     \caption{Properties of on-going and planned interferometric \HI\ wide  surveys}
          \centering
                   \label{tab:surveys}
         \begin{tabularx}{\textwidth}{l c c c c c c c c}
                          \hline\hline                   
         Survey  & max. z$^\ast$ & Area& D$^\dag$  &  Beam $^{\dag\dag}$& $\Delta v^{\ddagger}$ &noise$^\mathsection$ & \nhi$^\star$ & $\log$\mhi$^{\ddagger\ddagger}$ \\
	              &         &  [deg$^2$]  & [Mpc ]& ['', kpc at D]& [\kms]   & [mJy] & [$10^{18}$\cmsq]&  [\msun] \\
                  	\hline                                     
                    MeerKAT Fornax Survey & 0.58 & 12 &20 &21, 2.0 &1.4& 0.26 & 8.78&6.5 \\
                    MHONGOOSE & 0.58 & 30 & 10 &27, 1.3  &1.4& 0.15 & 4.46&5.7 \\
                    ViCTORIA & 0.58 & 104 &16 & 21 , 1.6 &11 & 0.65  &   40.4&6.7\\
                    MIGHTEE-HI & 0.58  & 32 & 270 &30, 35 &5.5 &0.10 & 4.82 &8.3\\
                    Apertif-Shallow  & 0.25  & 1000& 170 & 25, 23.4 &8.0& 1.6 &63.8  &9.1\\
                    Apertif-Medium Deep & 0.25 & 70&75 &25, 10.5  & 8.0& 0.51 & 22.3 &7.9\\ 
                    WALLABY & 0.26 & 14000 & 200& 30, 26.7 & 3.9& 1.6 &100 &9.3\\
                    FEASTS & 0.09 & 25 & 6.5 & 174,5.7 & 1.6 &0.96&0.54 & 6.1\\
                     \hline
                    FASHI & 0.09 & 22000 & 130 & 174,109 &6.9 &0.76&0.88 & 8.6\\
                    Euclid Deep Field South & 0.58 & 23 &270 & 30 , 35 &5.5 & 0.74  & 35.7  &9.2 \\                    
                    M+ Medium-shallow & 0.58 & 780 & 130 & 10, 6.05 & 5.5& 0.59  & 268  &8.5  \\
                    DSA2000-HI  (4 hrs) &1.0 &  $\delta \gtrsim -30^\circ$& 100 &3, 1.53  & 10 &0.098&190 &7.5\\
                    SKA-Mid (10 hrs, band 2,1)  & 0.5,3.0 &  $\delta \lesssim 30^\circ$ &  100& 4, 1.85 &5.5 & 0.037 &100 &7.0\\
                	\hline                   
         \end{tabularx} 
         \begin{itemize}
               \setlength{\itemsep}{1pt}
  \setlength{\parskip}{0pt}
  \setlength{\parsep}{0pt}
            \item[$^\ast$] Maximum accessible redshift by the survey in the L-band, except for the upcoming instruments DSA2000 and SKA-Mid.
            \item[$^\dag$] Median distance of the targets of the survey, or of their expected \HI\ detections.
            \item[$^{\dag\dag}$] Nominal angular resolution and spatial resolution at distance D of the surveys.
            \item[$^{\ddagger}$] Spectral resolution at the \HI\ rest frequency. For future MeerKAT surveys and SKA-Mid we assume a channel width of 26~kHz (5.5~\kms). 
            \item[$^\mathsection$] $1\sigma$ noise per channel with velocity width reported in column $\Delta v$. The values are taken from the respective survey papers and are not corrected for any differences in weighting or tapering.
            \item[$^\star$] $3\sigma$ column density sensitivity at the nominal spatial resolution of the surveys and common spectral resolution of $16$~\kms.
         \item[$^{\ddagger\ddagger}$] \HI\ mass sensitivity for a $3\sigma$, $50$~\kms\ unresolved source at the median distance of the survey. 
         \end{itemize}
         \label{Table1}
\end{table*}

\section{Acknowledgements}

This project has received funding from the European Research Council (ERC) under the European Union’s Horizon 2020 research and innovation programme (grant agreement no.~679627 and grant agreement no. 882793).

\end{document}